\documentclass[final,5p,times,twocolumn]{elsarticle}

\usepackage{amsthm}
\usepackage{dblfloatfix}
\usepackage{euscript}   % for the calligraphic G
\usepackage{graphicx}% Include figure files
\usepackage{dcolumn}% Align table columns on the decimal point
\usepackage{bm}% bold math
\usepackage{amsmath}% bold math
\usepackage{amssymb}

\usepackage{textcomp}%Provides extra symbols, e.g. arrows like \textrightarrow, various currencies (\texteuro,...), things like \textcelsius and many others.
\usepackage{tabularx} % Ruled Tabular
\usepackage{times}
\usepackage{titlesec}
\usepackage{wrapfig}
\usepackage{caption}
\usepackage{subcaption}
\usepackage{adjustbox}
\usepackage{listings}
\usepackage{silence}
\usepackage{wasysym}
\usepackage{titlesec}
\usepackage{graphicx}
\usepackage{colortbl}
\usepackage[table,xcdraw]{xcolor}
\usepackage{booktabs}

\usepackage{graphicx}% Include figure files
\usepackage[utf8]{inputenc}
\usepackage[T1]{fontenc}
\usepackage{etoolbox}
\usepackage{csquotes}
\usepackage{xcolor}

\usepackage{hyperref}% add hypertext capabilities
\hypersetup{
    colorlinks=true,
    linkcolor=green,
    filecolor=magenta,      
    urlcolor=cyan}

\usepackage{siunitx}
\usepackage{diagbox}
\begin{document}
\title{Temperature dependence of electronic conductivity from \textit{ab initio} thermal simulation}

\author[OU]{R. Hussein}
\ead{rh353321@ohio.edu}
\affiliation[OU]{organization={Department of Physics and Astronomy, Nanoscale and Quantum Phenomena Institute (NQPI)},
            addressline={Ohio University}, 
            city={Athens},
            postcode={45701}, 
            state={OH},
            country={USA}}

\author[LANL]{C. Ugwumadu}
\ead{cugwumadu@lanl.gov}
\affiliation[LANL]{organization = Quantum and Condensed Matter Physics (T-4) Group, addressline = { Los Alamos National Laboratory}, city = { Los Alamos}, postcode = {87545}, state ={NM}, country={USA}}

\author[OU]{K. Nepal}
%\ead{ kn478619@ohio.edu}

\author[LANL]{R. M. Tutchton}
%\ead{rtutchton@lanl.gov}

\author[PNNL]{K. Kappagantula}
%\ead{ keertisahithi.kappagantula@pnnl.gov}
\affiliation[PNNL]{organization={Pacific Northwest National Laboratory},
            % addressline={Ohio University}, 
            city={Richland},
            postcode={99354}, 
            state={WA},
            country={USA}}

\author[OU]{D. A. Drabold}
\ead{drabold@ohio.edu}

\date{}

\begin{abstract}
We present a temperature-dependent extension of the approximate electronic conductivity formula of Hindley and Mott that leverages time-averaged fluctuations of the electronic density of states obtained from \textit{ab initio} molecular dynamics. By thermally averaging the square of the density of states near the Fermi level, we obtain an estimate of the temperature dependence of the conductivity. This approach—termed the thermally-averaged Hindley-Mott (TAHM) method—was applied to five representative systems: crystalline aluminum (c-Al), aluminum with a grain boundary (Al\textsubscript{GB}), a four-layer graphene–aluminum composite (Al–Gr), amorphous silicon (a-Si) and amorphous germanium–antimony–telluride (a-GST). The method reproduces the expected Bloch–Grüneisen decrease in conductivity for c-Al and Al\textsubscript{GB}, even for temperatures well below the Debye temperature. Generally, the reduction (increase) in conductivity for metallic (semi-conducting) materials is reproduced. It captures microstructure-induced, thermally activated conduction in multilayer Al–Gr, a-Si and a-GST. Overall, the approach provides a computationally efficient link between time-dependent electronic structure and temperature-dependent transport, offering a simple and approximate tool for exploring electronic conductivity trends in complex and disordered materials.
\end{abstract}

 
\maketitle

\section{Introduction}

The electrical conductivity of condensed matter is a dynamical property that emerges from charge carriers moving through a time-dependent landscape influenced by lattice vibrations, defects, electronic scattering disorder, and external fields. Capturing that interplay—electrons responding to ionic motion on femto- to pico-second scales while also scattering from static inhomogeneities—is essential for predicting material performance and interpreting transport experiments across metals, semiconductors, and disordered materials. 

Two main approaches are used to compute electronic conductivity. The first is the semiclassical Boltzmann framework, in which charge carriers are treated as quasiparticles whose distribution obeys the Boltzmann transport equation (BTE). In the weak-field, near-equilibrium limit, the linearized BTE with a relaxation-time approximation (RTA)~\cite{Ziman2001,GiustinoRMP2017} yields a conductivity tensor and underpins most transport calculations in metals and semiconductors~\cite{Rocha2022,Claes2025}, as implemented, for example, in \textsc{BoltzTraP}~\cite{Madsen2006,Madsen2018} and \textsc{EPW}~\cite{Ponce2016}. Iterative solutions beyond the RTA can treat inelastic and anisotropic scattering more accurately~\cite{Ringhofer2002}, but the semiclassical picture still assumes coherent band transport and breaks down under strong disorder, localization, or ultrafast excitation.

When such effects become significant, fully quantum linear-response formulations are preferred. Within the quantum picture, conductivity may be obtained from many-body formulation of linear response theory~\cite{Kubo1957} or the single-electron formulation known as the Kubo–Greenwood formula (KGF), widely used with density functional theory (DFT)~\cite{Greenwood1958,Bose1993,Calderin2017,Dufty2018,Prasai2018,Subedi2021}.  

Mott writes the KGF for the conductivity as \cite{Mott1968,MottDavisBook1979}:
\begin{equation}\label{eq:mott}
    \sigma \sim \frac{2\pi e^2 \hbar^3}{m^2}\,\big |D(E_f)\big|^2_\mathrm{avg} N^2(E_F)
\end{equation}
where 
\begin{equation}
D(E_f) = \int \psi_{l^\prime}^* \tfrac{\partial}{\partial x} \psi_{l} d^3x
\end{equation}
for single-particle states $\psi$ and ``$\mathrm{avg}$" indicates an average over a small window near the Fermi level ($E_f$). $N(E_f)$ is the density of states energy at the Fermi level, and $\sigma$ is the DC conductivity.  From a Fermi Golden Rule argument it is natural to interpret electronic conduction in terms of quantum transitions at the Fermi level, leading to the conclusion that $\sigma \propto N^2(E_F)$. Hindley reached a similar conclusion by invoking a random phase approximation \cite{Hindley1970_RP}. We introduced what we call the "$N^2$ method" based upon this proportionality to obtain a positive additive distribution that provides information about the local intensity of electrical conductivity and applied it to a copper-carbon composite material \cite{NepalCarbon2025}, as well as defective tungsten \cite{UgwumaduPSSB2025}. These were obtained for the case of a static lattice. While one intuitively thinks of the electronic density of states (EDOS) at the Fermi level as a rough measure of conductivity or metallicity, in fact it is the \textit{squared} EDOS. 

The primary innovation of this paper is to exploit the $\sigma \propto N^2(E_F)$ approach by approximately including the effects of atomic motion -- estimating the temperature dependence of the electrical conductivity which is in general a challenging task. Abtew \textit{et al} \cite{abtew2007} and Subedi \textit{et al} \cite{subedi2022} showed that averaging the KGF over a suitably equilibrated \textit{ab initio} MD simulation (at constant temperature $T$) provides useful estimates for the $T$ dependence of $\sigma$, suggesting the possible utility of averaging $N^2$ in a similar way by estimating the temperature dependent conductivity as:
\begin{equation}\label{eq:sigma_t_mott}
    \sigma(T)\propto \frac{1}{n}\sum_{i=1}^n N^2(E_f, t_i)
\end{equation}
in which $i$ indexes a time step and the number of times steps $n$ is assumed to be large enough that the quantity $\sigma(T)$ is converged. In this paper, we showed that equation \ref{eq:sigma_t_mott} provides useful estimates of temperature-dependence and determine suitable run-times, ``$n$", post equilibration to obtain $\sigma(T)$ .

This approach is inherently approximate. Its main assumptions are: (1) an adiabatic treatment in which transport is estimated by averaging over "Born–Oppenheimer snapshots"; (2) classical dynamics, which neglects lattice quantization (phonons); (3) interpretation of the Kohn–Sham (KS) eigenvalues as a proxy for the electronic density of states at the Fermi level; (4) omission of any time-step dependence in the matrix element $D(E_f)$; and (5) applicability primarily to homogeneous systems—while the full Kubo–Greenwood formula yields a complete conductivity tensor, that tensor information is not retained in the present simplification. A practical advantage is that $N^2(E_f)$ can be extracted as a byproduct of quantum MD, making the workflow straightforward to implement.

In what follows, we show that this simple scheme, which we name the thermally averaged Hindley-Mott (TAHM) method, produces temperature trends for the conductivity comparable to those obtained from the more rigorous KGF method \cite{subedi2022} and from experiment \cite{chowdhury2013experimental}. Both the metallic reduction in conductivity with increasing temperature and semiconducting increase with increasing temperature is reproduced. As first seen by Subedi \textit{et al.} \cite{subedi2022}, the scheme produces reasonable results in a metal (Al) even for temperatures well below the Debye temperature, which is surprising for a classical simulation.

We also report a microstructure-dependent semiconducting behavior in aluminum–graphene composite with 4 layers of AB-stacked graphene featuring undulating (worm-like) morphology \cite{aG, aG2,NepalJphys2024}. Furthermore, we compare our results for amorphous silicon with KGF-based conductivity data \cite{abtew2007} as well as experimental conductivity data \cite{3_expt_diamond, black_expt, red_expt_liquid_1969}, and extend the analysis to the phase-change memory material, amorphous germanium–antimony–telluride \cite{NepalGST2026}.

\section{Methods}
\subsection{Structural Models}

%%%%%%%%% FIGURE 1 %%%%%%%%%%%%%%%%%%%%%%%%%%%%%%%%%
\begin{figure}[!tbph]
    \centering

     \includegraphics[width=\linewidth]{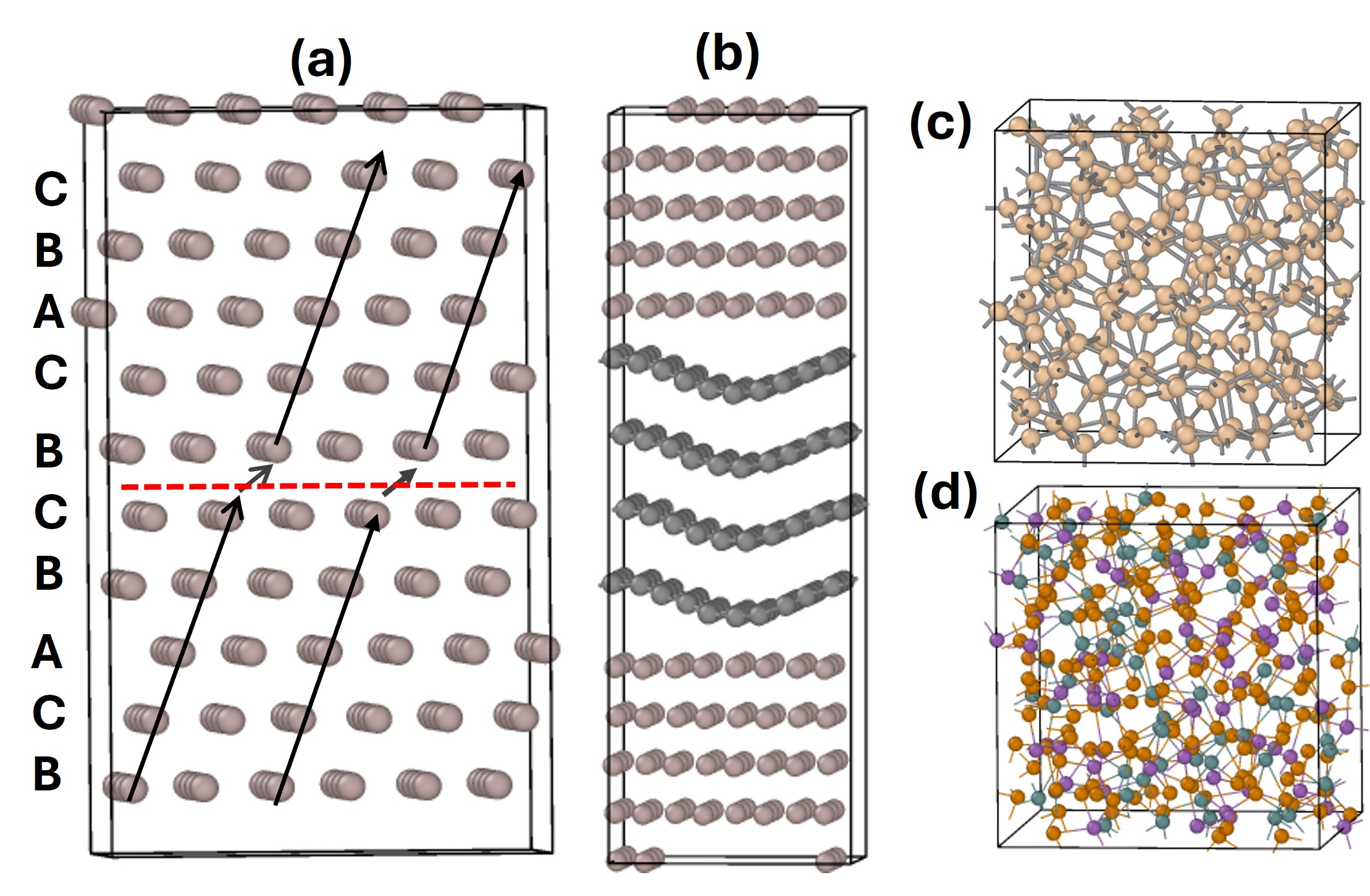}
    \caption{Structural representation of (a) aluminum with grain boundary (AB stacking fault represented by the red dashed line), (b) aluminum-graphene composite formed with the worm-like 4-layer (AB-stacked) graphene (gray), (c) amorphous silicon, and (d) amorphous germanium (teal)-antimony (purple)-telluride (brown) .}
    \label{fig:Fig_Models}
\end{figure}
%%%%%%%%% FIGURE 1 %%%%%%%%%%%%%%%%%%%%%%%%%%%%%%%%%

Five materials of interest are considered in this study: crystalline aluminum (c-Al), aluminum with a grain boundary (Al$_\text{GB}$), aluminum–graphene composites (Al–Gr), amorphous silicon (a-Si), amorphous germanium–antimony–telluride, Ge$_2$Sb$_2$Te$_5$ (a-GST).  The Vienna \textit{Ab-initio} Simulation Package (VASP)~\cite{VASP} was used for all simulations.

The c-Al model is a face-centered cubic (FCC) cubic containing 256 atoms (see Figure \textcolor{blue}{S1}e). This configuration is used for the bulk regions of the Al$_\mathrm{GB}$ structure, while the grain-boundary region contains a stacking fault (red dashed line in Figure~\ref{fig:Fig_Models}a). Similarly, the same c-Al configuration is used for the Al–Gr composite structure, in which the graphene region is a four-layer (AB-stacked) undulating (“worm-like”) sheet (Figure~\ref{fig:Fig_Models}b).

The undulating worm-like graphene morphology results from conjugate-gradient (CG) energy relaxation of the Al–Gr structure, starting from initially flat graphene layers between the Al slabs with an interfacial Al–C distance of 3.22~\AA. This configuration may mimic high-stress metastable interfaces produced by solid phase processing used to fabricate AA1100 alloys reinforced with reduced-graphene-oxide nanoparticles, yielding ultra-conductive Al composites~\cite{Nittala2023,NepalAPL2024}.

For a-Si, we employed the 216-atom model of Djordjevi\'c, Thorpe, and Wooten~\cite{DTW}, generated using the Wooten--Winer--Weaire approach~\cite{WWW} and previously analyzed in other studies \cite{IGRAM,UgwumaduSPTC2025} (Figure ~\ref{fig:Fig_Models}c). The 315-atom a-GST model (Figure~\ref{fig:Fig_Models}d) was taken from Reference~\cite{NepalGST2026} .

% \subsection{Temperature-dependent $\langle N^2 \rangle_t$ Method}
\subsection{Implementation}
The electronic density of states for all systems was computed using VASP \cite{VASP} with projector augmented-wave (PAW) potentials~\cite{PAW},  the Perdew–Burke–Ernzerhof (PBE) exchange–correlation functional~\cite{PBE}, and a Gaussian smearing width of 0.01 eV. For the time-dependent analysis, snapshots were extracted from \textit{ab initio} molecular dynamics (AIMD) simulations performed at different temperatures, maintained by a Nosé–Hoover thermostat~\cite{Nose}.

The AIMD trajectories spanned several picoseconds, with integration time steps of 0.5, 1.0, 0.75, 1.0, and 2.0~fs for c-Al (total simulation time = 2.5~ps), Al$_\text{GB}$ (5~ps), Al–Gr (3~ps), a-Si (4~ps), and a-GST (7.5-10~ps), respectively.  This framework is general and can equally incorporate electronic structures sampled from longer MD simulations based on classical, empirical, or machine-learning interatomic potentials. The $\Gamma$-point was used to sample the Brillouin zone for all simulations, which all employed periodic boundary conditions. The cutoff energy for a-GST was 320 eV and 400~eV was employed for the other structures. The selected simulation times were chosen to ensure convergence of the averaging at the Fermi level.

The instantaneous electronic density of states (EDOS) at time \(t_\mu\), (obtained from a constant temperature MD simulation), is:
\begin{equation}\label{eq:DoS_instanteneous}
    D(E,t_\mu) \propto \sum_i \delta\!\big(E - \epsilon_i(t_\mu)\big)
\end{equation}
 and \(\epsilon_i(t_\mu)\) is a Kohn--Sham eigenvalue at time step $t_\mu$. 

The Kohn-Sham states relevant to the DC conductivity are those near the Fermi level \(E_f\). We therefore define the instantaneous \(N^2\) at time \(t\) as:
\begin{align}
    N^2(E_f,T;t)
    &= \left[\int D(E,T,t)\,\delta_h(E - E_f)\,\mathrm dE \right]^2 \notag \\
    &=\left[\sum_i\delta_h(\epsilon_i(t)-E_f(t))\right]^2
    \label{eq:N2_t}\\[3pt]
    \delta_h(\varepsilon)
    &= \frac{1}{\sqrt{2\pi}\,h}\exp\!\left(-\frac{\varepsilon^2}{2h^2}\right)
    \label{eq:GaussianBroadening}
\end{align}
where we take \(\delta_h\) to be a Gaussian of width \(h\). The value of \(h\) is tuned to the presence (or absence) of an electronic gap: \(h = 0.35~\mathrm{eV}\) for c-Al and Al\(_\text{GB}\), \(0.2~\mathrm{eV}\) for Al--Gr, \(0.7~\mathrm{eV}\) for a-Si, and \(0.54~\mathrm{eV}\) for a-GST. These values are larger than the mean spacing between KS eigenvalues at \(E_f\), so that \(h\) is broad enough to capture thermal fluctuations of near-\(E_f\) states while remaining narrow enough to resolve intrinsic spectral features.\footnote{For small systems or when few states lie near \(E_f\), this broadening choice becomes more delicate.} At each time step, the Fermi level is shifted so that \(E_f=0\) for reference.

To proceed, we require a long trajectory of a well-equilibrated atomic configuration at temperature $T$ (see Equation~\ref{eq:sigma_t_mott}). We sample the MD trajectory using a running (cumulative) time average over post-equilibration time steps $\{t_k\}^K_{k=1}$. Using Equations \ref{eq:mott}--\ref{eq:GaussianBroadening}, we compute TAHM ($\langle N^2 \rangle_t$) as:
\begin{equation}
  \langle N^2 \rangle_t
  \;=\;
  \frac{1}{K}
  \sum_{k = 1}^{K}
  N^2(E_f,T;t_k)
\end{equation}
This yields an estimate of $\sigma(T)$ once fluctuations becomes small for a sufficiently long simulation.

\subsection{Estimating Conductivity from $\langle N^2 \rangle_t$}\label{sec:fittingN2}
The temperature dependence of the electronic conductivity can be estimated by correlating $\langle N^2 \rangle_t$ with experimental conductivity data at a known temperature. Fitting a single data point of $\langle N^2 \rangle_t$ to a corresponding experimental conductivity provides an estimate of the proportionality constant associated with the conductivity matrix element \cite{NepalCarbon2025}.

For a given material $\mathrm{m}$, with known experimental conductivity at a given temperature, $T_0$ as $\sigma^\mathrm{m}_\mathrm{exp}(T_0)$. We first define a single calibration factor \(\eta^m\) at \(T_0\) from the measured conductivity as:

\begin{equation}
  \eta^\mathrm{m}
  = \left.
  \frac{\sigma^\mathrm{m}_\mathrm{exp}}
       {\langle N^{2} \rangle_t^{\mathrm{m}}}\right|_{T_0}
    \label{eq:eta_def}
\end{equation}
We then use this proportionality to map the TAHM of the material, $\mathrm{m}$, onto a predicted conductivity and resistivity at any temperature, $T$, as:
\begin{equation}
  \sigma^\mathrm{m}_{\mathrm{N^2}}(T)
  = \eta^\mathrm{m}\,\big\langle N^{2}(T)\big\rangle_t^\mathrm{m} \;;
  \qquad
  \rho_{\mathrm{N^2}}^\mathrm{m}(T)
  = \frac{1}{\eta^\mathrm{m}\,\big\langle N^{2}(T)\big\rangle_t^\mathrm{m}}
  \label{eq:calibrated_sigma}
\end{equation}
hence giving a quantitative estimate of temperature-dependent electronic transport behavior. We summarize the $\eta$ value for all systems in Table \ref{tab:calibration}. We note that no calibration was performed for a-Si as we opted to show its trend instead, due to a jump in conductivity at high temperatures, as discussed in Section \ref{sec:a-Si}.

%%%%%%%%%%%%%% TABLE FOR Experimental Calibraiton %%%%%%%%%%%%%%%
\begin{table}[t!]
	\caption{Experimental calibration summary for the materials. No calibration was done for amorphous silicon (a-Si).}
	\label{tab:calibration}
	\begin{tabular*}{\linewidth}{@{\extracolsep\fill}cccc}
        Material & T [K] & $\sigma_\mathrm{exp}^\mathrm{m}$~[S/m] & $\eta^\mathrm{m}$\\
        \hline
    c-Al / Al$_\mathrm{GB}$   & 50 & 1.81$\times10^9$~\cite{chowdhury2013experimental} & 7.59$\times10^{-6}$\\
    % Al$_\mathrm{GB}$ & 50 & 1.81$\times10^9$~\cite{chowdhury2013experimental} & 7.59 $\times10^{-6}$\\
    Al-Gr   & 300 & 3.87$\times10^7$~\cite{Smyrak2025} & 1.49$\times10^4$\\
    a-Si & --- & --- & ---\\
    a-GST & 300 & 2.22~\cite{Kato2005} & 5.51$\times10^{-4}$\\
	\hline
		\end{tabular*}
    
\end{table}
%%%%%%%%%%%%%% END TABLE FOR NPT Models %%%%%%%%%%%%%%%

\begin{figure*}[!tbhp]
    \centering
    \includegraphics[width=.9\linewidth]{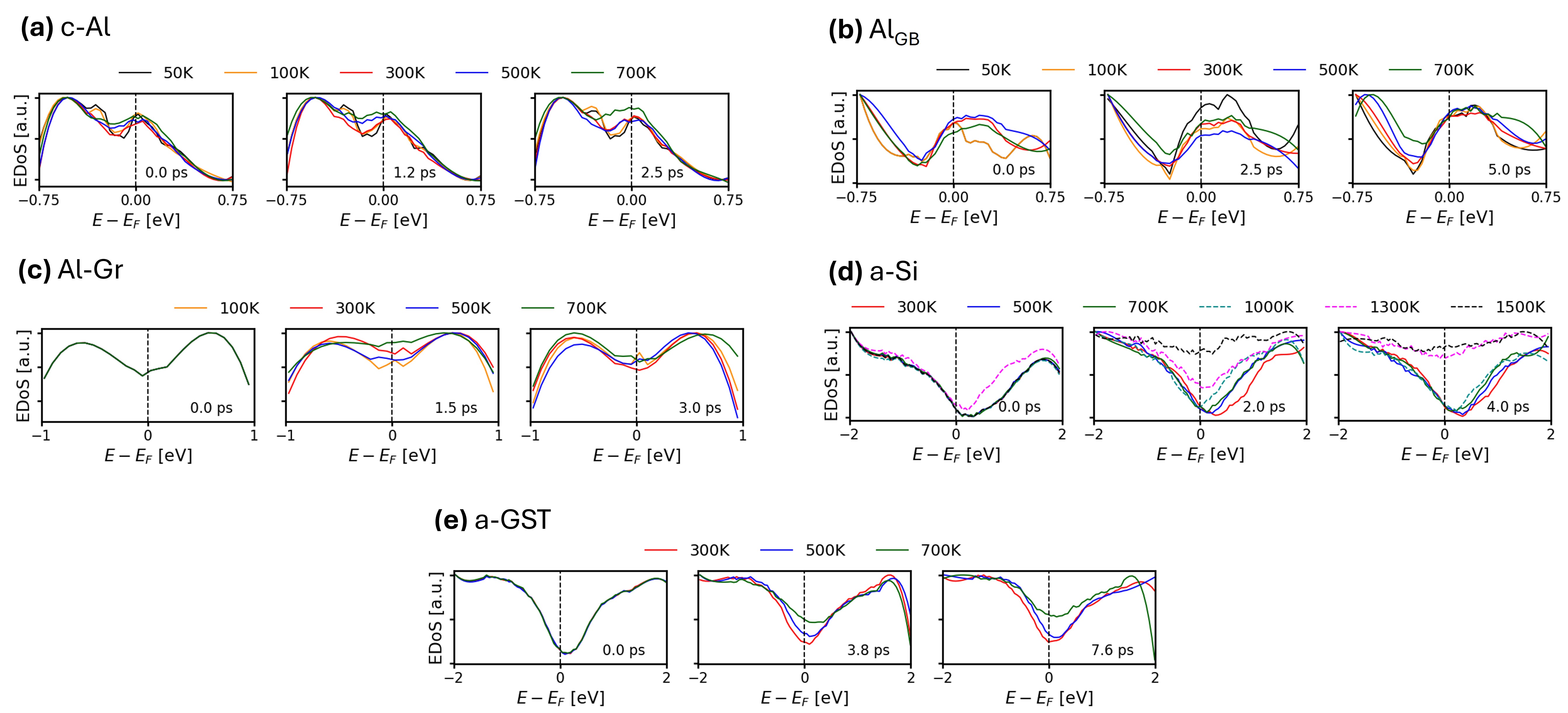}
    \caption{The electronic density of states near the Fermi level obtained from the MD simulations at different instantaneous Born-Oppenheimer snapshots, showing for  (a) Crystalline aluminum, (b) Aluminum with a grain boundary, (c) Aluminum-graphene composite, (d) amorphous silicon, and relaxed (e) amorphous germanium–antimony–telluride.}
    \label{fig:Fig_EDOS}
\end{figure*}

\section{Results and Discussion}
For brevity, we will denote $N^2(E_f,T,t)$ in Equation \ref{eq:N2_t} by $N_t^2$. The character of temperature-induced EDOS fluctuations around the Fermi level differs among the materials. The instantaneous EDOS at selected time steps and temperatures for all structures is shown in Figure \ref{fig:Fig_EDOS}a--e, and all similar plots for all temperatures considered per structure is shown in Figure~\textcolor{blue}{S2}a--e. Except for a-Si, which was analyzed over the broader range of 200--1800~K, the temperature range for the other systems extends up to 700~K. For clarity, the Fermi level in each plot has been shifted to zero (indicated by the dashed black line).

Appreciable temperature-dependent fluctuations near $E_F$ are observed for c-Al, Al$_\text{GB}$, and Al–Gr (Figure \ref{fig:Fig_EDOS}a--c). The fluctuations are most pronounced for a-Si (Figure \ref{fig:Fig_EDOS}d), where the electronic gap broadens with increasing temperature and simulation time. They are less pronounced for a-GST (Figure \ref{fig:Fig_EDOS}e). The following subsections present a detailed analysis for each system.

%%%%%%%%% FIGURE 3 %%%%%%%%%%%%%%%%%%%%%%%%%%%%%%%%%
\begin{figure}[!t]
    \centering
    \includegraphics[width=.87\linewidth]{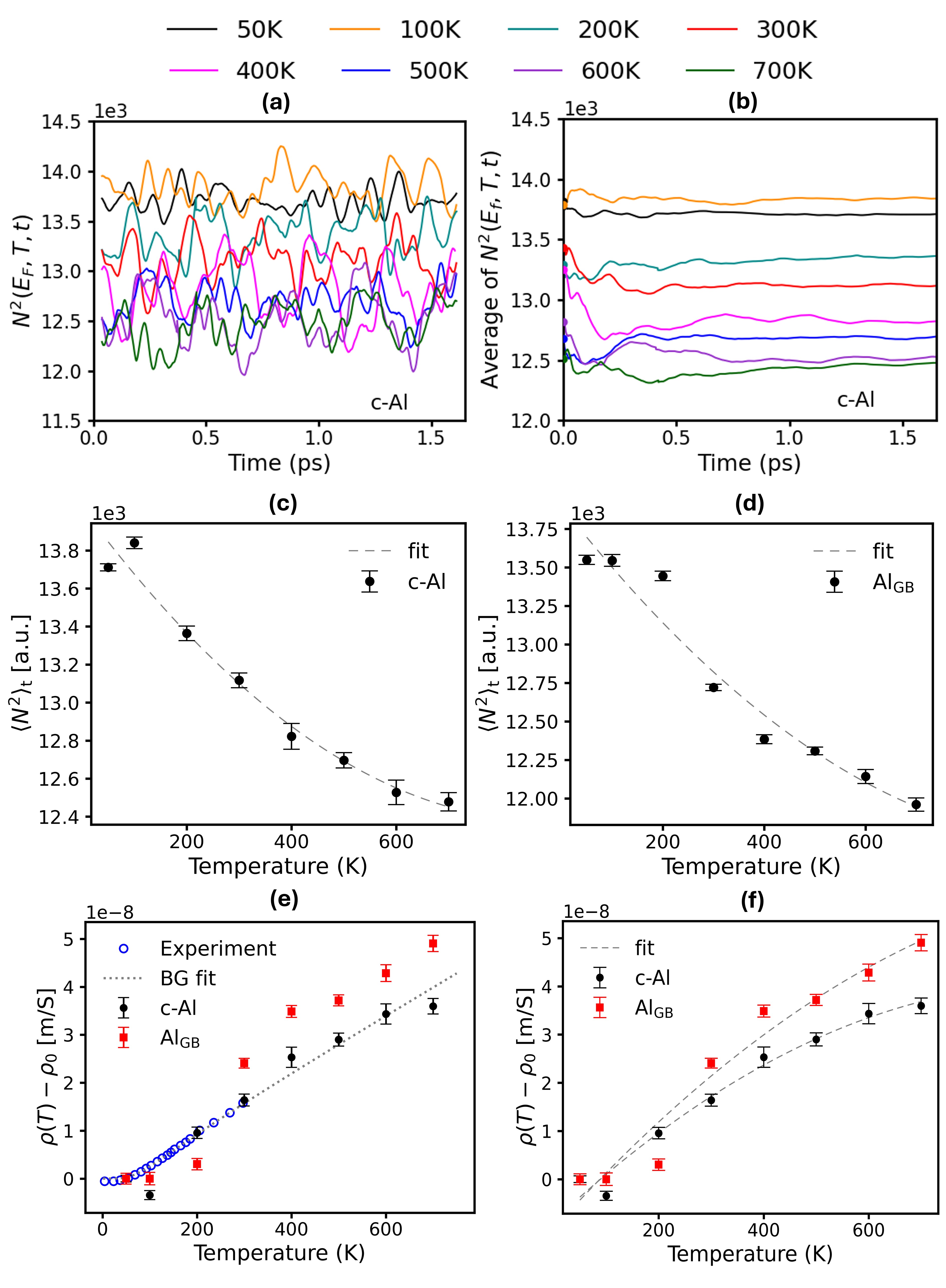}
    \caption{Analysis for crystalline aluminum (c-Al) and aluminum with a grain boundary (Al$_\text{GB}$) . 
 (a) Instantaneous $N_t^2$ from the EDOS at different temperatures and (b) the convergence of the running time-average of $N_t^2$ for c-Al (similar plots for Al$_\text{GB}$ are provided in Figure \textcolor{blue}{S3}a and b). Converged $\langle N^2 \rangle_t$ versus temperature with quadratic fit for (c) c-Al and (d) Al$_\text{GB}$.  (e) Experimental resistivity~\cite{chowdhury2013experimental} compared with values from $N^2(E_F)$ for c-Al and Al$_\text{GB}$; Bloch–Grüneisen (BG) predictions are included.  (f) Temperature-dependent resistivity for c-Al and Al$_\text{GB}$; from $\langle N^2 \rangle_t$, with error bars from MD time averaging. }
    \label{fig:Fig_Al}
\end{figure}
%%%%%%%%% FIGURE 3 %%%%%%%%%%%%%%%%%%%%%%%%%%%%%%%%%

\subsection{Aluminum: Crystalline and Grain Boundary Structures}
The analysis for the c-Al and Al$_\text{GB}$ systems were performed over a temperature range of 50--700~K. The characteristic electronic density of states (EDOS) profile of aluminum is largely preserved, exhibiting extended near-$E_f$ states, in both c-Al (Figure~\ref{fig:Fig_Al}a,b) and Al$_\text{GB}$ (Figure~\textcolor{blue}{S3}a,b). Early MD steps that exhibit strong non-equilibrium fluctuations before thermal stabilization were not included in the analysis. For c-Al (Al$_\text{GB}$), this included the first 0.8~ps (2 ps). The running average implementation on $N^2_t$ effectively smoothened these oscillations to yield convergence as shown in Figure~\ref{fig:Fig_Al}b and ~\textcolor{blue}{S3}b for c-Al and Al$_\text{GB}$, respectively.  The time-averaged $\langle N^2 \rangle_t$ were determined once the running average varied by less than 5\% over successive steps for 1 ps, and exhibit a decrease from 100~K to 700~K in both c-Al (Figure~\ref{fig:Fig_Al}c) and Al$_\text{GB}$ (Figure~\ref{fig:Fig_Al}d). 

Next, we compare $\langle N^2 \rangle_t$ results with experiment in Figure~\ref{fig:Fig_Al}e. The experimentally measured resistivity from Reference \cite{chowdhury2013experimental} (up to 300~K) is plotted against the resistivity inferred from the $\langle N^2 \rangle_t$ using Equation \ref{eq:calibrated_sigma}. To convert $\langle N^2 \rangle_t$ into a conductivity estimate, we choose a reference temperature $T_0 = 50$~K and determine a proportionality constant $\eta$  to be $\approx 7.59 \times 10^{-6}$ (Table \ref{tab:calibration}). This choice is consistent with the KGF analysis of Subedi \textit{et al.}~\cite{subedi2022}, where they also adopt $T_0=50$~K for the same dataset.   For crystalline aluminum, electron–phonon scattering follows the Bloch–Grüneisen (BG) behavior which suggests that in the low-$T$ regime ($T \ll \Theta_{\mathrm{D}}$) the phonon contribution scales steeply ($\propto T^5$) and becomes small compared to the defect-induced residual resistivity~\cite{Ashcroft1976solid}.%\textcolor{red}{There is a factor of 5$\times10^{-3}$ for the resistivity plot shown.} 

For c-Al in Figure \ref{fig:Fig_Al}e, the temperature dependence predicted by $\langle N^2 \rangle_t$ is consistent with the KGF analysis \cite{subedi2022} and agrees qualitatively with both the experimental data and the extrapolated BG fit. The resistivity of Al$_\text{GB}$ is higher than that of c-Al and shows a larger deviation from experiment, especially at high temperature, as expected due to the additional grain-boundary scattering.  Nevertheless, at lower temperatures its resistivity remains close to c-Al and to the experimental curve, and the overall trend with $T$ is preserved, as seen more clearly in Figure~\ref{fig:Fig_Al}f. Between 200 and 700~K, crystalline Al shows a $\approx 73\%$ reduction in conductivity, compared to $\approx 94\%$ for Al$_\text{GB}$. This behavior parallels the SPC-based results of Reference~\cite{subedi2022}, where local conductivity is reduced at Al grain boundaries.

\subsection{Aluminum–Graphene Composite}
\begin{figure}[!tpbh]
    \centering
    \includegraphics[width=\linewidth]{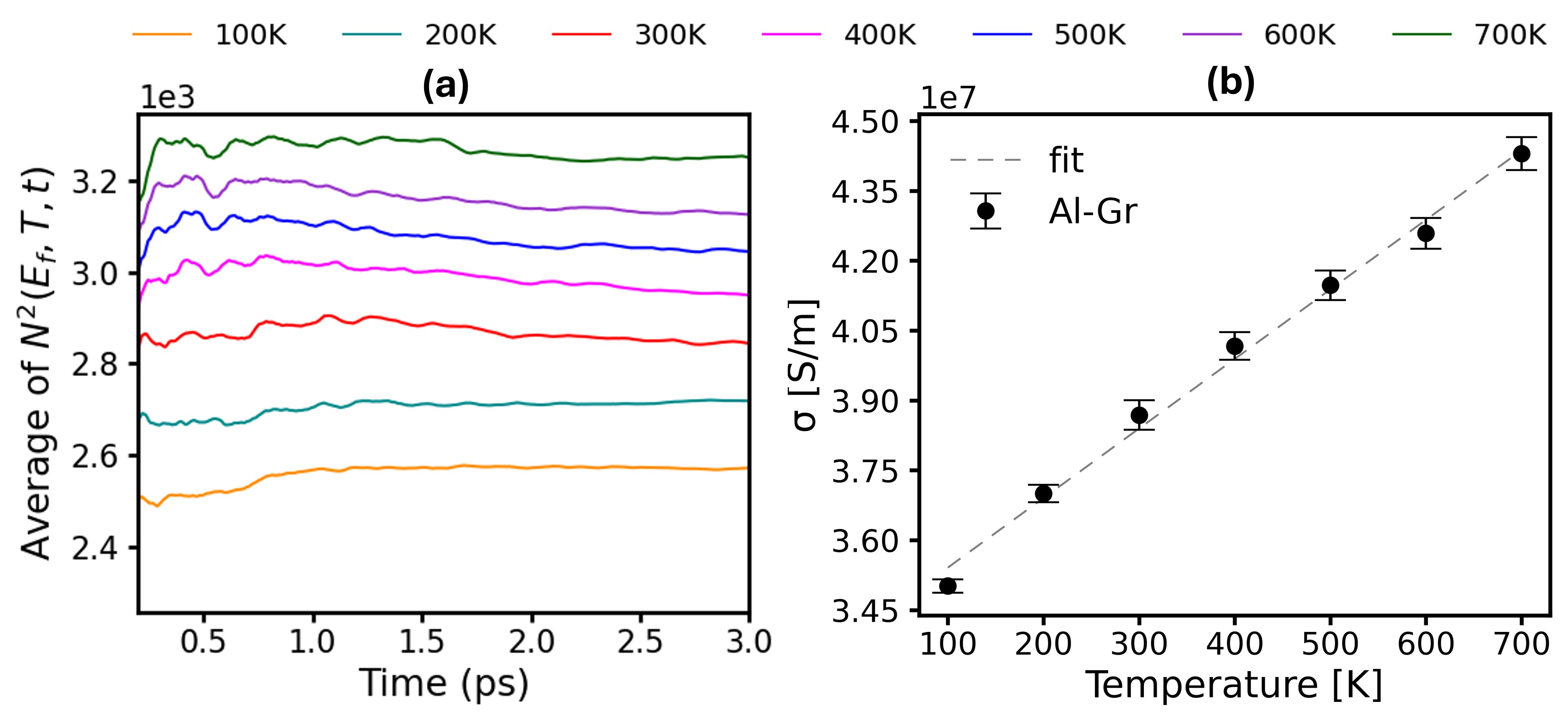}
    \caption{Analysis for aluminum-graphene composite structure.The running time-average of $N_t^2$ illustrating convergence.(b) Temperature dependent $\langle N^2 \rangle_t$ conductivity extrapolated from experimental conductivity at 300 K from Reference \cite{Smyrak2025}.}
    \label{fig:Fig_Al_Gr}
\end{figure}

The running time average of $N_t^2$ for Al–Gr converges well and increases smoothly with temperature over the range 100--700~K (Figure~\ref{fig:Fig_Al_Gr}a). Instantaneous $N_t^2$ traces at the simulated temperatures are shown in Figure~\textcolor{blue}{S4}. Following $\eta$ calibration at 300~K (see Table \ref{tab:calibration}), we found that $\sigma_{\mathrm{N^2}}(T)$ exhibits an approximately linear increase with temperature, with slope $1.49\times10^{4}~\mathrm{S\,m^{-1}\,K^{-1}}$ (Figure~\ref{fig:Fig_Al_Gr}b). This trend indicates a semiconducting-like response at the Al–Gr interface where thermal fluctuations promote near-$E_F$ states and enhance the time-averaged overlap between occupied and unoccupied electronic states.

This behavior contrasts with previous simulations of flat single- and double-layer Al–Gr composites, which report a metal-like decrease in $\sigma(T)$ with increasing temperature~\cite{NepalAPL2024}, although at fixed temperature the conductivity increases as the Al--C distance is reduced from 3.41 to 2.97~\AA.  The metallic-to-semiconducting behavior in the Al–Gr composite arises from its microstructure---aluminum coupled to a multilayer, undulating graphene stack---which appears to stabilize and thermally activate interfacial conduction channels, leading to the monotonic increase in conductivity. To our knowledge, this provides the first atomistic evidence of semiconducting behavior in multilayer ($>3$-layer) Al–Gr composites with a worm-like graphene morphology, consistent with high-resolution imaging observations by Kappagantula \textit{et al.} \cite{KK_observation}.

%%%%%%%%% FIGURE 4 %%%%%%%%%%%%%%%%%%%%%%%%%%%%%%%%%
%%%%%%%%% FIGURE 4 %%%%%%%%%%%%%%%%%%%%%%%%%%%%%%%%%

\begin{figure}[!tbhp]
    \centering
    \includegraphics[width=\linewidth]{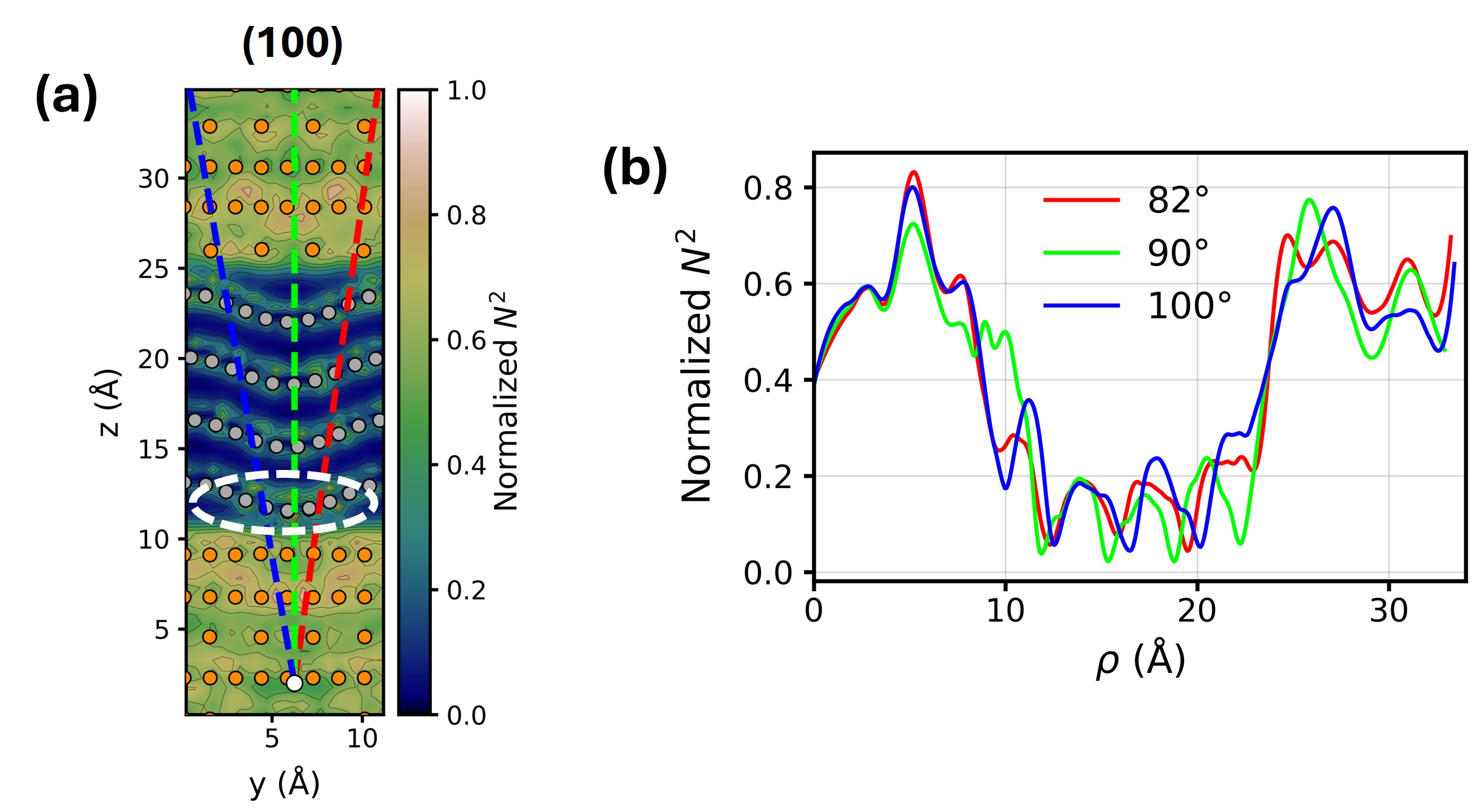}
    \caption{ Spatial projection of $N^2$ electronic activity in the Al-graphene composite. (a) 2D colormap of normalized $N^2$ on a (100) plane slice at $x \approx 6.0$~\AA. (b) Radial ($\rho$) profiles of $N^2$ extracted along the colored traces in (a) at the indicated projection angles $\theta$.}
    \label{fig:Fig_Al_Gr_N2}
\end{figure}

The $N^2$ method \cite{NepalCarbon2025} was employed to visualize conduction-active regions in the aluminum--graphene composite. Whereas TAHM captures the time-averaged near-$E_f$ electronic activity along an MD trajectory, $N^2$ projects the near-$E_f$ contributions into real space, revealing where conduction pathways form across heterogeneous regions. Figure~\ref{fig:Fig_Al_Gr_N2}a shows the (100) cross-section at $x = 6\,\text{\AA}$. The dashed white oval marks the region where the lower carbon layer (gray spheres) is closest to the upper aluminum matrix (orange spheres), with a distance of $\approx 2.36\,\text{\AA}$. The contour map (normalized $N^2$) spans 0 (blue) to 1 (white). The colored traces (green, red, and blue) are radial cuts taken along near-by in-plane directions ($82^\circ$, $90^\circ$ and  $100^\circ$), showing how near-$E_f$ activities varies with distance from the Al/C interface. The resulting peak-suppression-recovery pattern in Figure~\ref{fig:Fig_Al_Gr_N2}b reflects enhanced interfacial electronic activity, its attenuation into the amorphous graphene layers, and its re-emergence beyond the Al-C interaction zone.

\subsection{Amorphous and Liquid Silicon}\label{sec:a-Si}

\begin{figure}[!thpb]
    \centering
    \includegraphics[width=\linewidth]{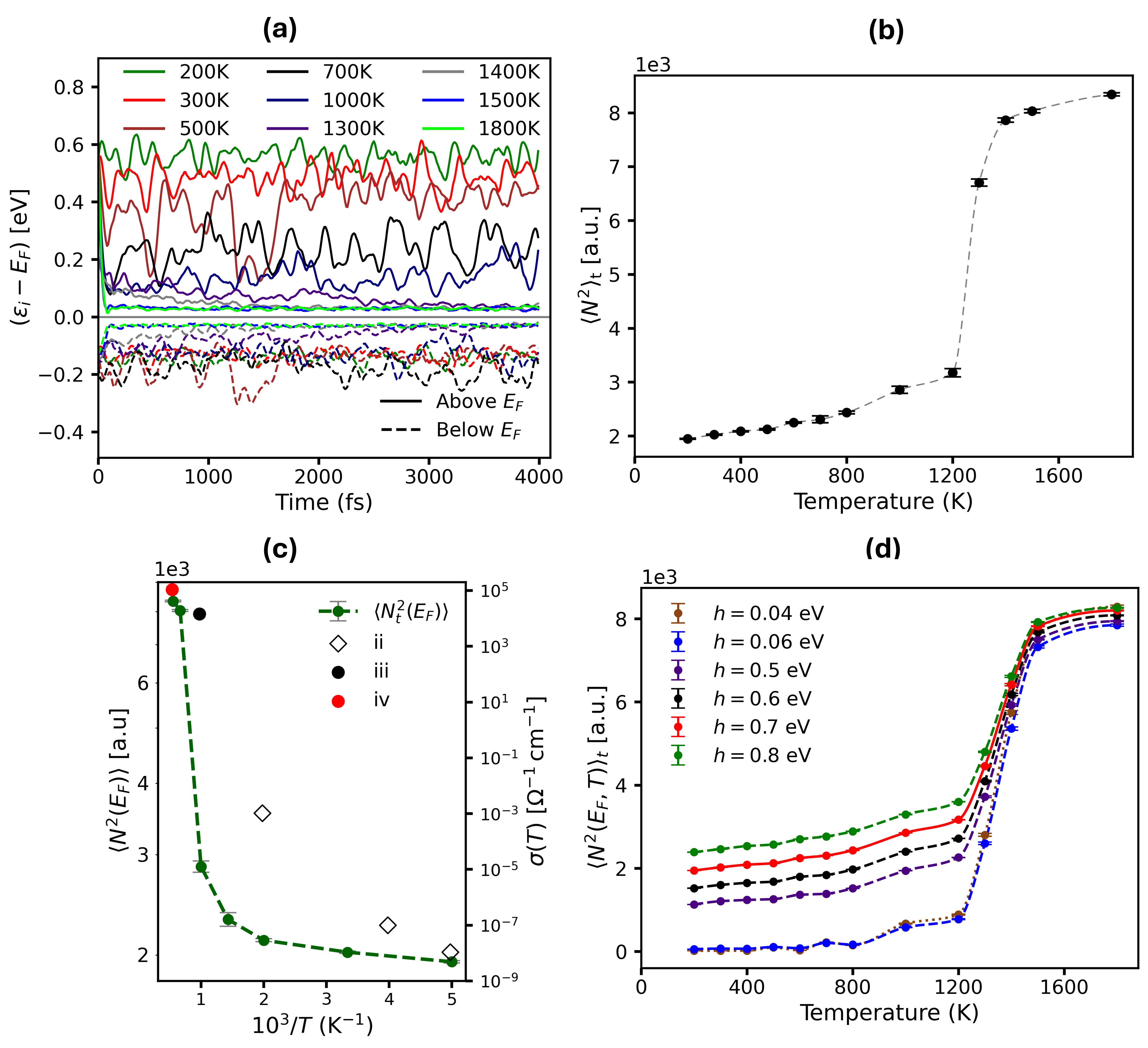}
    \caption{Analysis for amorphous silicon. (a) Thermal fluctuation of near-gap states ($E_f = 0$) at temperatures between $200$ to $1800$K. (b) Converged $\langle N^2 \rangle_t$ values as a function of temperature. (c) Comparison of $\langle N^2 \rangle_t$ with experimental data points from References (ii)~\cite{3_expt_diamond}, (iii)~\cite{black_expt} and (iv)~\cite{red_expt_liquid_1969}. (d) Dependence of  $\langle N^2 \rangle_t$ on the Gaussian broadening ($h$) used to obtain the instantaneous $N^2(E_f,T,t)$.} 
    \label{fig:Fig_aSi}
\end{figure}

Figure~\ref{fig:Fig_aSi}a shows the temporal fluctuations of the mean Kohn-Sham eigenvalues of the states near the Fermi level at each temperature, serving as a direct probe of near-$E_F$ electronic fluctuations induced by thermal lattice motion for temperatures from 200 to 1800~K. At low temperatures, the EDOS above and below $E_F$ remain well separated, persisting the band-gap. With increasing temperature, particularly beyond 1200~K (and around its melting point of $\approx$ 1420~K~\cite{Donovan1983}), pronounced broadening and overlap of these states are observed, signaling thermally driven delocalization of the electronic states near the band edges \cite{abtew2007, Drabold1991, Li2003}. Figure \textcolor{blue}{S5}a and b shows the instantaneous $N_t^2$ across the temperatures and its  running time-average convergence.

%%%%%%%%% FIGURE 5 %%%%%%%%%%%%%%%%%%%%%%%%%%%%%%%%%
%%%%%%%%% FIGURE 5 %%%%%%%%%%%%%%%%%%%%%%%%%%%%%%%%%

$\langle N^2 \rangle_t$ remains nearly constant up to about 1000~K and then increases sharply between 1200 and 1500~K (Figure~\ref{fig:Fig_aSi}b) . This rapid rise, followed by a slight plateau, mirrors the experimentally observed behavior of the electrical conductivity of liquid silicon at high temperatures~\cite{Sasaki1995}. The trend reflects a phase transition as the system approaches its melting point ($\approx$ 1420~K;  Figure~\ref{fig:Fig_EDOS}d). At these elevated temperatures, strong thermal atomic motion and dynamic structural arrangements in the disordered liquid network enhance the overlap among electronic states near the Fermi level, facilitating carrier hopping and improving electronic connectivity \cite{abtew2007}.

We remind the reader that $\eta$ calibration was not carried out for a-Si due to the melting phase transition at about 1420 K. Figure \ref{fig:Fig_aSi}c compares the trend of $\langle N^2 \rangle_t$ to experimentally measured conductivities~\cite{3_expt_diamond, black_expt, red_expt_liquid_1969}. Both show the characteristic exponential rise of conductivity with temperature expected for thermally activated transport in amorphous semiconductors, followed by a transition toward non-semiconducting behavior near the melting regime. The agreement between the computed and experimental trends indicates $\langle N^2 \rangle_t$ captures the essential physics of thermally assisted delocalization in a-Si \cite{Li2002,Li2003}.

We also investigate the effects of different Gaussian broadening, $h$ used to obtain $N^2(E_f,T,t)$ in Equation \ref{eq:N2_t}.  As shown in Figure~\ref {fig:Fig_aSi}d, the overall trend remains consistent for different $h$ values from 0.04 eV to 0.8 eV; with wider spread at temperatures below the jump at 1200 K, and more convergence at higher temperatures (> 1350 K). 

\subsection{Germanium–Antimony–Telluride}
\begin{figure}[!tpbh]
    \centering
    \includegraphics[width=\linewidth]{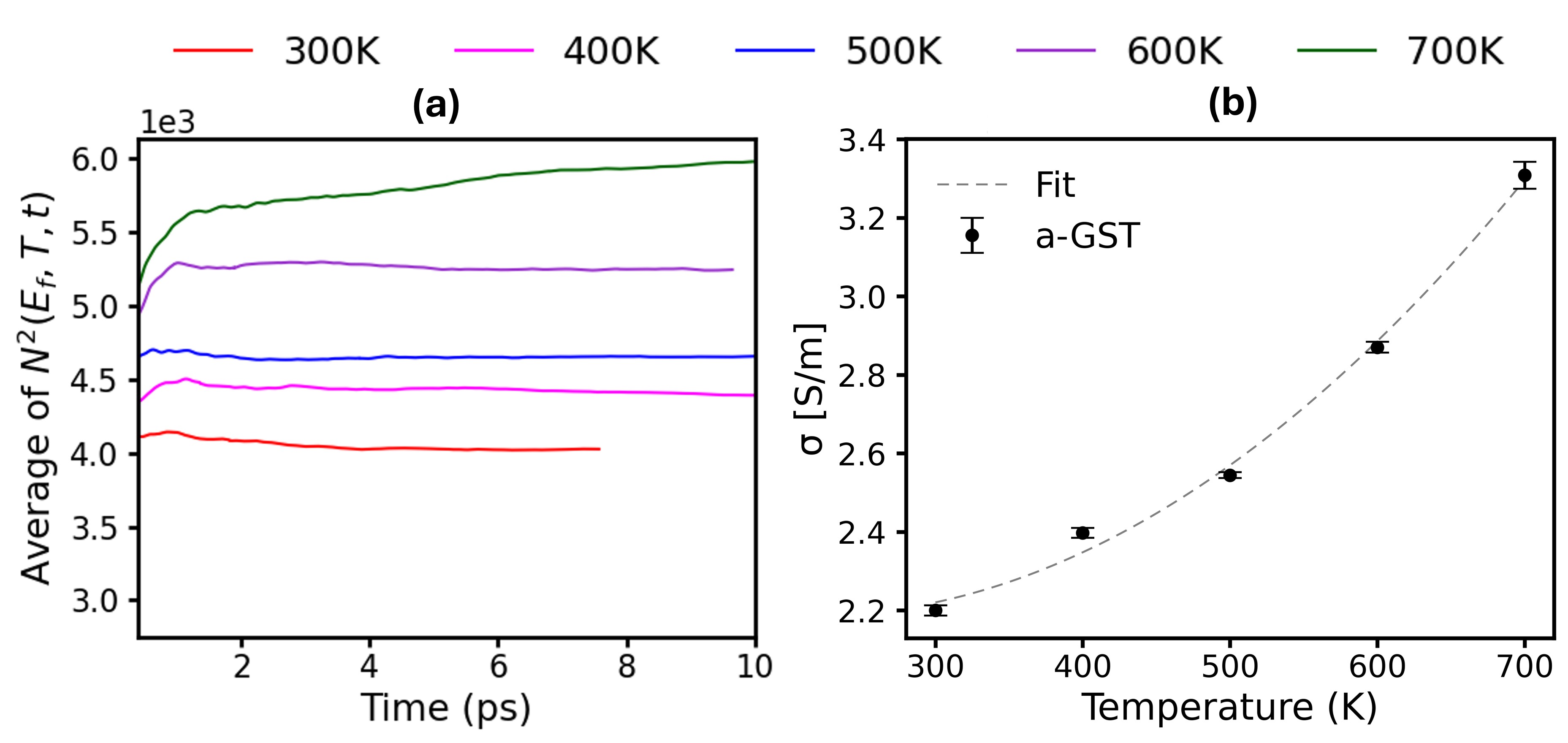}
    \caption{Analysis for amorphous Germanium–Antimony–Telluride. (a) The running time-average of $N_t^2$ illustrating convergence. (b) Temperature dependent $\langle N^2 \rangle_t$ conductivity extrapolated from experimental conductivity at 300 K from Reference ~\cite{Kato2005}.}
    \label{fig:Fig_aGST}
\end{figure}

Similar to the Al–Gr composite, a-GST exhibits semiconducting behavior characterized by an increase in $\langle N^2 \rangle_t$ with temperature as shown by the convergence of the running time-average plot in Figure \ref{fig:Fig_aGST}a for the temperature range of 300 -- 700 K (See the instantaneous $N_t^2$ plot in Figure \textcolor{blue}{S6}). This trend is consistent with the expected response of a-GST, where thermal excitation promotes carrier activation across its narrow mobility gap~\cite{Lee2005,Muneer2018}. The a-GST model employed in this study exhibits a band-gap of approximately 0.54 eV, with a mid-gap state located about 0.31~eV below the LUMO (See model ``M5" in Reference \cite{NepalGST2026}). 

 %%%%%%%% FIGURE 6 %%%%%%%%%%%%%%%%%%%%%%%%%%%%%%%%%
%%%%%%%%% FIGURE 6 %%%%%%%%%%%%%%%%%%%%%%%%%%%%%%%%%

The conductivity of a-GST, shown in Figure \ref{fig:Fig_aGST}b shows nearly linear increase with temperature. This is is consistent with experimental reports of thermally activated conductivity in a-GST~\cite{Muneer2018,Kato2005}. The $\langle N^2 \rangle_t$ fit to conductivity was carried out using the experimentally measured average conductivity of a-GST at 300~K ($\sigma_\mathrm{exp}^\mathrm{GST} \approx 2.22$~S/m) gives a proportionality constant of $\eta \approx 5.51\times10^{-4}$ (Table \ref{tab:calibration}).

\section{Conclusion}
We have developed and demonstrated the TAHM method that extends Mott and Hindley's simplified picture of electronic transport into the time domain. By averaging the squared fluctuations of near-Fermi-level electronic density of states obtained along \textit{ab initio} molecular dynamics trajectories, the method captures the coupling between lattice motion, electronic disorder, and charge transport. When scaled to a single experimental conductivity value, the resulting temperature-dependent trends reproduce the observed behavior across metallic, semi-conducting, composite, and amorphous systems. The method predicts metallic reduction in conductivity as well as semiconducting increase with increasing temperature from TAHM.

For FCC aluminum and aluminum with a grain boundary, TAHM shows a monotonic decrease with increasing temperature, consistent with Bloch–Grüneisen electron–phonon scattering. In contrast, the multilayer, worm-like aluminum–graphene composite displays a semiconducting-like increase in TAHM with temperature, reflecting thermally activated interfacial conduction pathways stabilized by its microstructure. In amorphous silicon, TAHM remains nearly constant at low temperature but rises sharply between 1200 and 1500~K, coinciding with the onset of gap closure and the semiconductor-metal (melting) transition. The a-GST system exhibits a steady, nearly linear increase in TAHM with temperature, consistent with thermally activated carrier excitation across a mobility gap.

Together, these results establish the TAHM method as a simple yet predictive microscopic descriptor of temperature-dependent electronic transport that is transferable across diverse structural and electronic regimes. The framework provides a computationally efficient and physically transparent complement to other Kubo–Greenwood-based formulations, enabling rapid assessment of conductivity trends in nanostructured, composite, and disordered materials.

% acknowledgment
% \section*{Acknowledgement} 
% We gratefully acknowledge Dr. X for valuable discussions.

% Funding
\section*{Funding}
\noindent C.U. and R.M.T. acknowledge support from the Laboratory Directed Research and Development (LDRD) program at Los Alamos National Laboratory (LANL) through the Director’s Postdoctoral Fellowship Program (Project No. 20240877PRD4). LANL is operated by Triad National Security, LLC, for the U.S. Department of Energy (DOE) National Nuclear Security Administration under Contract No. 89233218CNA000001. C.U. and R.M.T. also acknowledge additional support from the U.S. DOE, Office of Science, National Quantum Information Science Research Centers, and Quantum Science Center. K. N.  acknowledges financial support from the Nanoscale \& Quantum Phenomena Institute (NQPI), conferred through the NQPI graduate research fellowship. K.K. and D.A.D. acknowledge support from the U.S. DOE Advanced Materials and Manufacturing Technologies Office (AMMTO) through the CABLE Program, through the Pacific Northwest National Laboratory (PNNL). PNNL is operated by the Battelle Memorial Institute for the U.S. DOE under Contract No. DE-AC0676RL01830.  D.A.D. acknowledges support from the the U.S. National Science Foundation (Project No. MRI-2320493) for computational resource, and from the U.S. Office of Naval Research (Project No. N000142312773).

\bibliographystyle{elsarticle-num}
\bibliography{TAHM_ref}
\end{document}